\documentclass[12pt]{article}
\usepackage{amsmath}
\usepackage{graphicx}
\newcommand{\al}[1]{\alpha_{#1}}
\newcommand{\expp}{e^{\frac{i}{2} p \times p'}}
\newcommand{\intalp}{\int_0^\infty d\al{1} d\al{2} d\al{3}}
\newcommand{\inta}{\int_0^1 d\al{1} d\al{2} d\al{3} \delta(1-\sum\al{i})}
\newcommand{\T}{\al{1}\mg^2+(\al{2}+\al{3})^2m^2-\al{2}\al{3}q^2}
\newcommand{\dir}[1]{\not\!{#1}}
\newcommand{\N}{m^2(\al{1}-\al{2}-\al{3})+\mg^2(\al{2}+\al{3})+m^2(\al{2}+\al{3})^2-\al{2}\al{3}q^2}
\newcommand{\aaalp}{\al{1}+\al{2}+\al{3}}
\newcommand{\aalp}{\al{2}+\al{3}}
\newcommand{\mg}{m_{\gamma}}
\newcommand {\newsection}{\setcounter{equation}{0}\section}
\begin{document}

\thispagestyle{empty}

\begin{flushright}
IC/2000/XX \\
hep-th/0008132
\vskip 14mm
\end{flushright}

\begin{center}
{\LARGE{ Noncommutative QED and Anomalous Dipole Moments}}
\vskip 1cm

{\textbf {\Large {Ihab. F. Riad and M.M. Sheikh-Jabbari}}}\footnote
{A Dissertation Presented by I.F.R. to the ICTP High Energy Section in Candidacy for
the Diploma Degree, August 2000}

\vspace{12 mm}

{\it The Abdus Salam International Centre for Theoretical Physics\\
 Strada Costiera, 11. 34014, Trieste, Italy}\\
{\tt  ifriad, jabbari@ictp.trieste.it }\\

\end{center}

\vskip 8mm

\begin{center}
{\textbf Abstract}
\end{center}

We study QED on noncommutative spaces, NCQED. In particular we present the detailed
calculation for the noncommutative electron-photon vertex and show that the Ward
identity is satisfied. We discuss that in the noncommutative case moving electron
will show {\it electric} dipole effects. In addition, we work out the electric and
magnetic dipole moments up to one loop level. For the magnetic moment we show that 
noncommutative electron has an intrinsic (spin independent) magnetic moment.

\vskip 4cm

\newpage
%\input{ack.tex}

%\newpage 
\baselineskip 0.8 cm

\tableofcontents
\vskip 1cm

\baselineskip 0.65 cm

\newsection{Introduction}

Although, physics on the noncommutative spaces has a long standing story \cite{snyder},
recently it has been re-motivated by string theory arguments \cite{CDS}. Apart from the string
theory interests the field theories on such spaces, noncommutative field theories (NCFT's), in
their own turn are very interesting. 
Although being non-local, it has been argued that they can be treated
as sensible field theories, and in the last two years there have been a  
lot of work devoted to the study of NCFT's.  The question of renormalizability of NCFT's in general and in
particular NC scalar and NC Yang-Mills (NCYM) theories have been addressed extensively
\cite{{NCYM},{aref},{seib},{Suss},{Haya},{Armoni},{Chiral},{Alvarez1},{Alvarez2},{Micu}}
\footnote{ Because of the huge number of papers on this issue, hereby we apologize
for all the related works which have not been quoted.}.

For the NC scalar theories it has been shown that real $\Phi^4$ theory in 4 dimensions is two loop 
renormalizable \cite{{aref},{Micu}}. The pure NCYM have been studied only in one loop level and shown
to be renormalizable. The $\beta$-function of these theories (pure NCYM) is found to be the same as
the corresponding commutative theory. 
  
The problem of adding fermions (matter fields) has not been studied in detail. In particular,
noncommutative version of QED, NCQED, has been discussed in
\cite{{Haya},{Chiral},{Alvarez1},{CPT},{Roiban}}.
We emphasize that here we only consider  spaces-like noncommutativities and not the time-like one
(noncommutative space-times).
In the latter case, it has been shown that the corresponding field theories are not unitary \cite{Mehen}.
However, for the light-like noncommutativity it has been shown that we still have a well-defined
quantum theory \cite{Light-like}.

In this work we study some details of NCQED. Working out the electron-photon 
interaction vertex in NCQED up to one loop, we find the so-called vertex functions
and thereby we read off the anomalous magnetic moment. By explicit calculation we
show that Ward identity is satisfied in the NCQED case. Since we have an
extra suitable vector in our theory, there is some room for new type of magnetic
moment which is spin independent and is proportional to the noncommutativity
parameter.

As it has been discussed in \cite{{Alvarez1},{Dip}} particles in the noncommutative  
gauge theories show {\it electric dipole} effects. 
We recall this property in the classical case, and
also study one loop  quantum corrections to the electric dipole moment.

The paper is organized as following. In section 2, we briefly review the preliminaries we need; we
introduce noncommutative spaces and field theories in general and specify the NCQED by presenting its 
classical action and basic Feynman graphs. Then in section 3, we work out details of noncommutative 
electron-photon vertex at one loop level. In addition, renormalizing the
corresponding graphs in the minimal subtraction (MS) scheme, we discuss the IR/UV 
mixing which is a characteristic of any NCFT.  
In section 4, using the renormalized vertex functions we find
"electromagnetic dipole" properties of noncommutative electron. We end this work by conclusions and
remarks.

\section{Preliminaries}
%%%%%%%%%%%%%%%%%%%%%%%%%%%%%%%%%%%%%%%%%%%%%%%%%%%%%%%%%%%%%%%%%%%%%
 
\subsection{Noncommutative spaces} 

Usual quantum mechanics is formulated on commutative spaces
satisfying the following commutation relations, 
\begin{eqnarray} [\hat{X}_i,\hat{P}_j] & = &
i\hbar\delta_{ij} ~~~~~~~~\textrm{and}, \nonumber \\ {}[\hat{X}_i,\hat{X}_j]  = 0\ \
\ \ & , &\;\;\;\;\;
[\hat{P}_i,\hat{P}_j] = 0.  
\end{eqnarray} 
Then in order to describe a noncommutative space,
the above commutation relations should be changed as, 
\begin{eqnarray} [\hat{X}_i,\hat{P}_j] & = &
i\hbar\delta_{ij}, \nonumber \\ {}[\hat{X}_i,\hat{X}_j] & = & i\theta_{ij} ~~~~~~~~~~~~\textrm{and}
\nonumber \\ {}[\hat{P}_i,\hat{P}_j] & = & 0\ , 
\end{eqnarray} 
where $\theta_{ij}$, is the noncommutative
{\it constant} of dimension $[M]^{-2}$. We can see that for theories on such
spaces the Lorentz symmetry is explicitly violated however,
it will be recovered in the  $\theta \to 0$ limit. The above
noncommutative theory can be expanded to include noncommuting space-time, i.e.  
\begin{equation}
[\hat{X}_\mu,\hat{X}_\nu] = i\theta_{\mu \nu}.  
\end{equation} 
These type of theories were studied in
\cite{Mehen} and  shown to suffer from the loss of unitarity, so for the purpose of
our work here, we will only consider the noncommuting spaces.  Since our field
theory is better formulated through the path integral formulation, 
we can implement the noncommutivity of space into path
integral formulation through what is known as the Weyl-Moyal correspondence
\cite{Alvarez2}\footnote{In the following equations there are some obvious factors
of $2\pi$ which we avoid them here.} ,

$$
\hat{\Phi}(\hat{X}) \longleftrightarrow \Phi(x)\ ; 
$$

\begin{eqnarray} 
\hat{\Phi}(\hat{X}) & = & \int_\alpha e^{i\alpha \hat{X}} \phi(\alpha) d\alpha,
\nonumber \\ \phi(\alpha) & = & \int e^{-i\alpha x} \Phi(x) dx, 
\end{eqnarray} 
where $\alpha$ and $x$
are real variables. Then, 
\begin{eqnarray} \hat{\Phi}_1(\hat{X})\hat{\Phi}_2(\hat{X}) & = &
\iint_{\alpha \beta} e^{i\alpha \hat{X}} \phi(\alpha) e^{i \beta \hat{X}} \phi(\beta) d\alpha d\beta
\nonumber \\ & = & \iint_{\alpha \beta} e^{i(\alpha+\beta)\hat{X}-\frac{1}{2} \alpha_\mu
\beta_\nu[\hat{X}_\mu ,\hat{X}_\nu]} \phi(\alpha) \phi(\beta) d\alpha d\beta, 
\end{eqnarray} 
hence
\begin{equation} 
\hat{\Phi}_1(\hat{X})\hat{\Phi}_2(\hat{X}) \longleftrightarrow
\left(\Phi_1 * \Phi_2\right)(x), 
\end{equation} 

\begin{equation} \label{star} 
\left(\Phi_1 * \Phi_2\right)(x) \equiv
\left[e^{\frac{i}{2}\theta_{\mu \nu} \partial_{\zeta \mu} \partial_{\eta \nu}} \Phi(x+\zeta)
\Phi(x+\eta)\right]_{\zeta=\eta=0}.  
\end{equation} 

In other words, the noncommutative version of a field theory is obtained by
replacing all the field products by the star product (\ref{star}). 
It is easy to check that the Moyal bracket of coordinates defined by, 
\begin{equation}
[x_\mu,x_\nu]_{MB} = x_\mu * x_\nu - x_\nu * x_\mu\ , 
\end{equation} 
satisfies the commutation relations on the noncommutative spaces (2.3).

\subsection{The Noncommutative QED}

In this part we introduce the structure of the action for NCQED \cite{Haya}. 
We will notice that due to the presence of the star product and the Moyal brackets, 
the noncommutative U(1) is similar to non-Abelian gauge theories. As usual, the action for
a gauge theory consists of two parts, the gauge fields and
matter fields, fermions. To start with, we  write the term for the gauge fields:

\begin{equation}\label{lym}
S_{YM}  =  -\int d^4x \frac{1}{4e^2}F_{\mu \nu}*F^{\mu \nu}\ ,
\end{equation}
with
\begin{equation}
F_{\mu \nu}  =  \partial_\mu A_\nu - \partial_\nu A_\mu -i[A_\mu,A_\nu]_{MB} \ .
\end{equation}
The above action enjoys the noncommutative U(1) symmetry, defined by 
\begin{eqnarray}
A_\mu \to A'_\mu(x) & = & U(x)*A_\mu*U(x)^{-1}+iU(x)*\partial_\mu U(x)^{-1} \ , \\
U(x)&=& exp*(i\lambda) \equiv  1+i\lambda-\frac{1}{2}\lambda*\lambda\ + \ .... \ .
\end{eqnarray}
Under the above transformation, the field strength $F_{\mu \nu}$, transforms
as
\begin{equation}
F_{\mu \nu} \to F_{\mu \nu}' = U(x)*F_{\mu \nu}*U(x)^{-1}\ ,
\end{equation} 
and hence
\begin{equation}
S_{YM}\rightarrow S'_{YM}  =  -\int d^4x \frac{1}{4e^2}U(x)*F_{\mu
\nu}*U(x)^{-1}*U(x)*F^{\mu \nu}*U(x)^{-1} \ .
\end{equation}
using, 
%\begin{equation}
$$
U(x)*U(x)^{-1}  =  U(x)^{-1}*U(x) = 1 \ ,
$$
%\end{equation} 
and the cyclic property of the star product under the integral (see appendix \ref{A}), 
we conclude that (\ref{lym}) is invariant under the above defined gauge
transformation. 
Also using the property that under the integral the star product is commuting, one can remove 
the star product between two $F$'s, then 
\begin{eqnarray}
S_{YM} & = & -\int d^4x \frac{1}{4e^2}F_{\mu \nu}F^{\mu \nu}\ .
\end{eqnarray} 

In order to add the matter fields, we need the notion of
covariant derivative, $D_\mu$, for which we demand that our fermionic 
sector is also invariant under the above mentioned noncommutative local gauge
transformations. We notice that one can extend the local gauge transformations to
fermions in the following two different ways,
\begin{equation}
\Psi(x) \to \Psi'(x)  =  U(x)*\Psi(x) \ ,
\end{equation}
or
\begin{equation}
\tilde{\Psi}(x) \to \tilde{\Psi'}(x)  =  \tilde{\Psi}*U(x)^{-1}\ .
\end{equation}

Therefore, one can define two covariant derivatives corresponding to two different
fermions:
\begin{equation}
D_\mu\Psi  =  \partial_\mu\Psi - iA_\mu*\Psi\ ,
\end{equation}
or
\begin{equation}
D_\mu\tilde{\Psi}  =  \partial_\mu\tilde{\Psi} + i\tilde{\Psi}*A_\mu \ .
\end{equation}
From the above two equations we see that by taking the limit, 
$\theta_{\mu \nu} \to 0$, the star product disappears and the 
two fermion fields $\Psi(x)$ and $\tilde{\Psi}(x)$ will show different charges.
In fact it has been shown that these two fermions are related by the noncommutative
version of the charge conjugation \cite{CPT}.

So, altogether the full Lagrangian can be written as,
\begin{eqnarray}
S & = & \int d^4x \left(-\frac{1}{4e^2}F_{\mu\nu}F^{\mu\nu}+\bar{\Psi}(\not\!\!{D}-m)\Psi + 
L_{gauge} + L_{ghost}\right)\ ,
\end{eqnarray}   
where we have also added the term $L_{gauge}$ for gauge fixing. Since the quadratic terms are
the same as in the usual QED, we have the same gauge fixing term. 
The $L_{ghost}$ accounts for the ghost fields appearing in the gauge fixing procedure, 
similar to that of the non-Abelian gauge theories. 

%%%%%%%%%%%%%%%%%%%%%%%%%%%%%%%%%%%%%%%%%%%%%%%%%%%%%%%%%%%
%%%%%%%%%%%%%%%%%%%%%%%%%%%%%%%%%%%%%%%%%%%%%%%%%%%%%%%

\subsection{Feynman rules for NCQED}
Using the property of the star product under the integral sign and that, the quadratic terms are not 
changed in the presence of the star product, we conclude that the propagators 
for the free fermion, gauge and the ghost fields  are
the same as in the case of usual QED, i.e. 

\begin{equation}
\includegraphics[bbllx=70,bblly=400,bburx=600,bbury=500,width=2cm]{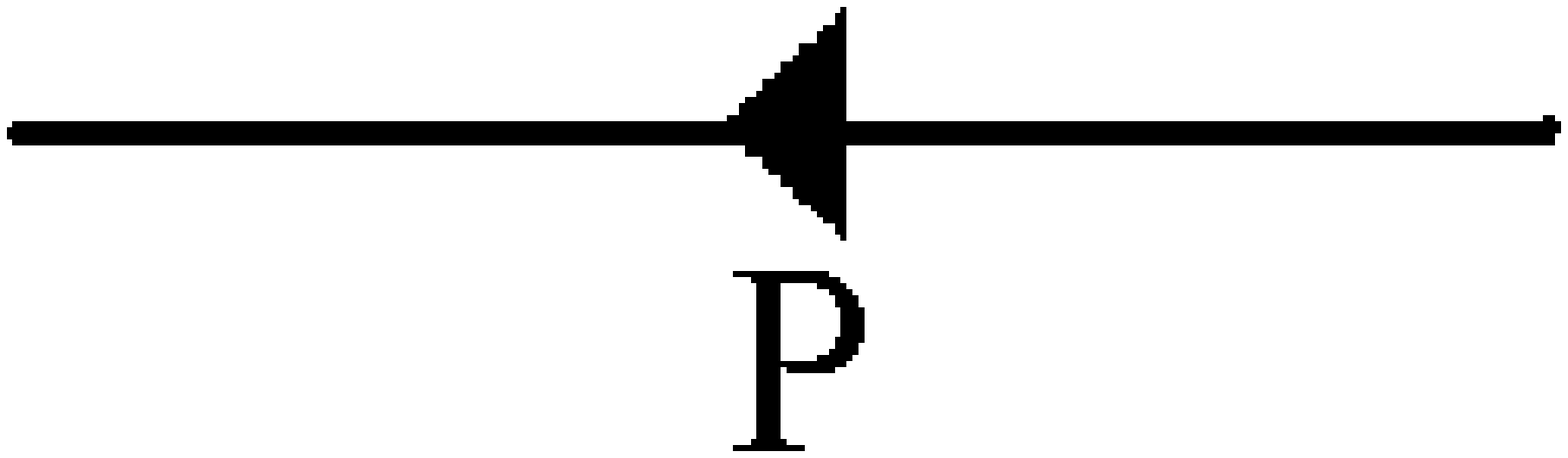}
= \frac{i}{\dir{p}-m+i\epsilon}\ ,
\end{equation}

\begin{equation}
\includegraphics[bbllx=70,bblly=400,bburx=600,bbury=500,width=2cm]{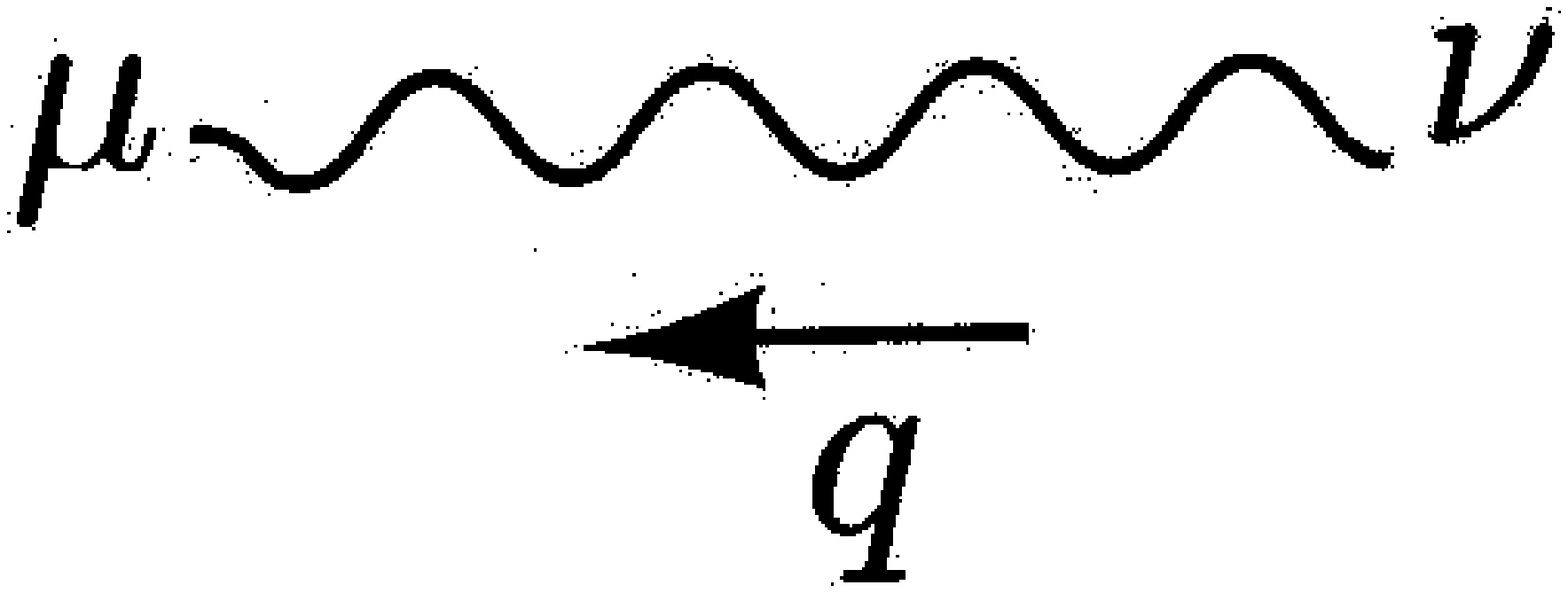}
= \frac{g^{\mu\nu}}{i(p^2+i\epsilon)}\ ,
\end{equation}

\begin{equation}
\includegraphics[bbllx=150,bblly=700,bburx=250,bbury=500,width=2cm]{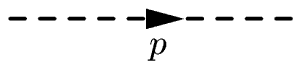}
= \frac{-1}{i(p^2+i\epsilon)}\ ,
\end{equation}
where we were using the Feynman gauge.

For the interaction terms, we see that they are similar to those of
non-Abelian gauge theories \cite{NCYM,Haya},
in the sense that we get cubic and quadric interaction 
vertices for the gauge fields besides the usual 
vertices found usual QED. Here we just present the Feynman rules, to show their similarities
and differences with the non-Abelian case.

\begin{eqnarray}
\includegraphics[bbllx=1,bblly=700,bburx=250,bbury=500,width=6cm]{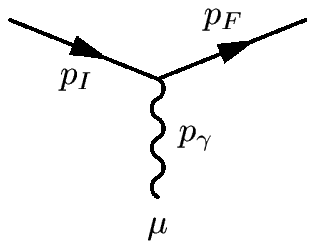} & = &  
ie\gamma^\mu e^{\frac{i}{2}p_I \times p_F} \nonumber \\
\nonumber \\
\nonumber \\
\nonumber \\
\includegraphics[bbllx=1,bblly=730,bburx=250,bbury=500,width=7cm]{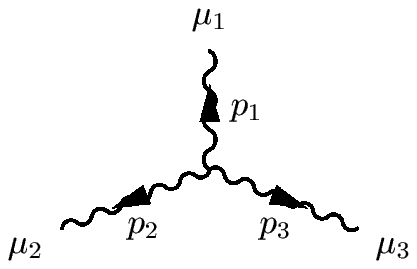} & = &  
-2e \sin\left(\frac{1}{2}p_1\times p_2\right) \nonumber \\
& & \times\Big[(p_1-p_2)^{\mu_3}g^{\mu_1 \mu_2}  \nonumber \\
& & +(p_2-p_3)^{\mu_1}g^{\mu_2 \mu_3}  \nonumber \\
& & +(p_3-p_1)^{\mu_2}g^{\mu_3 \mu_1}\Big] \nonumber \\
\nonumber \\
\nonumber \\
\includegraphics[bbllx=10,bblly=730,bburx=290,bbury=500,width=7cm]{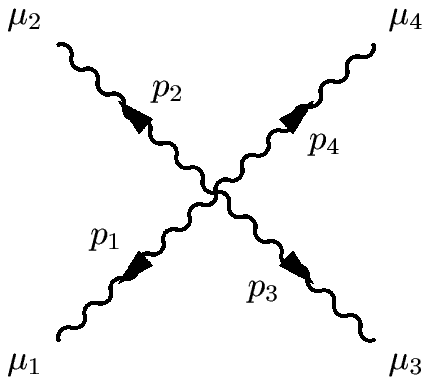} & = & 
-4ie^2\Big[\left(g^{\mu_1 \mu_3}g^{\mu_2 \mu_4} - g^{\mu_1 \mu_4}g^{\mu_2 \mu_3}\right) \nonumber \\
& & \times \sin\left(\frac{1}{2}p_1 \times p_2\right)\sin\left(\frac{1}{2}p_3 \times p_4\right) \nonumber \\
& &\Big[\left(g^{\mu_1 \mu_4}g^{\mu_2 \mu_3} - g^{\mu_1 \mu_2}g^{\mu_3 \mu_4}\right) \nonumber \\
& & \times \sin\left(\frac{1}{2}p_3 \times p_1\right)\sin\left(\frac{1}{2}p_2 \times p_4\right) \nonumber \\
& & \Big[\left(g^{\mu_1 \mu_2}g^{\mu_3 \mu_4} - g^{\mu_1 \mu_3}g^{\mu_2 \mu_4}\right) \nonumber \\
& & \times \sin\left(\frac{1}{2}p_1 \times p_4\right)\sin\left(\frac{1}{2}p_2 \times p_3\right) \nonumber \\  
\nonumber \\
\nonumber \\
\includegraphics[bbllx=1,bblly=680,bburx=250,bbury=500,width=7cm]{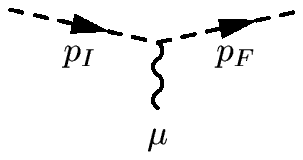} & = & 
2iep_F^\mu \sin(\frac{1}{2}p_I \times p_F) \ . 
\end{eqnarray}

\vspace{1cm}
We observe that all vertices here are similar to those in non-Abelian gauge theories
in which  the structure constant is replaced by $2sin(\frac{1}{2}p\times p')$.
This can be seen if we notice that the structure constants appear because of the commutation 
$[A_a,A_b]=if_{abc}A^c$ in non-Abelian theories. Hence we expect the appearance of the
factor $2sin(\frac{1}{2}p\times p')$ as a consequence of the Moyal bracket, i.e.

\begin{eqnarray}
[A_\mu,A_\nu]_{MB} & = & A_\mu*A_\nu-A_\nu*A_\mu \nonumber \\
& = & \int\ d^4p\ d^4p'\  A_\mu(p) A_\nu(p') \bigg(e^{\frac{i}{2}p\times
p'}-e^{\frac{-i}{2}p\times p'}\bigg)e^{i(p+p').x}\nonumber \\
& = & \int d^4p\ d^4p'\ 2iA_\mu(p) A_\nu(p') \sin(\frac{1}{2}p\times p')e^{i(p+p').x}\ .
\end{eqnarray}

\newsection{Electron-photon vertex at one loop level}

 In the previous section we introduced the NCQED, and showed that in this theory there are new
type of vertices similar to those found in non-Abelian gauge theories. In this section we
perform explicit calculation of the vertex function for the photon-electron at the one loop level 
which are expected to contribute to the anomalous magnetic moment.

%%%%%%%%%%%%%%%%%%%%%%%%%%%%%%%%%%%%%%%%%%%%%%%%%%%%%%%%%%
\subsection{Vertex structure at the one loop level}

In the case of NCQED and because of the three photon vertices, the 
electron-photon vertex receives contributions from the two diagrams of figure.1 .

%%%%%%%%%%%%%%%%%%%%%%%%%%%%%%%%%%%%%%%%%%%%%%%%%%%%%%%%%%%%%%%%%%%%%%%%%

\begin{figure}[h]
\begin{center}
\includegraphics[scale=.5]{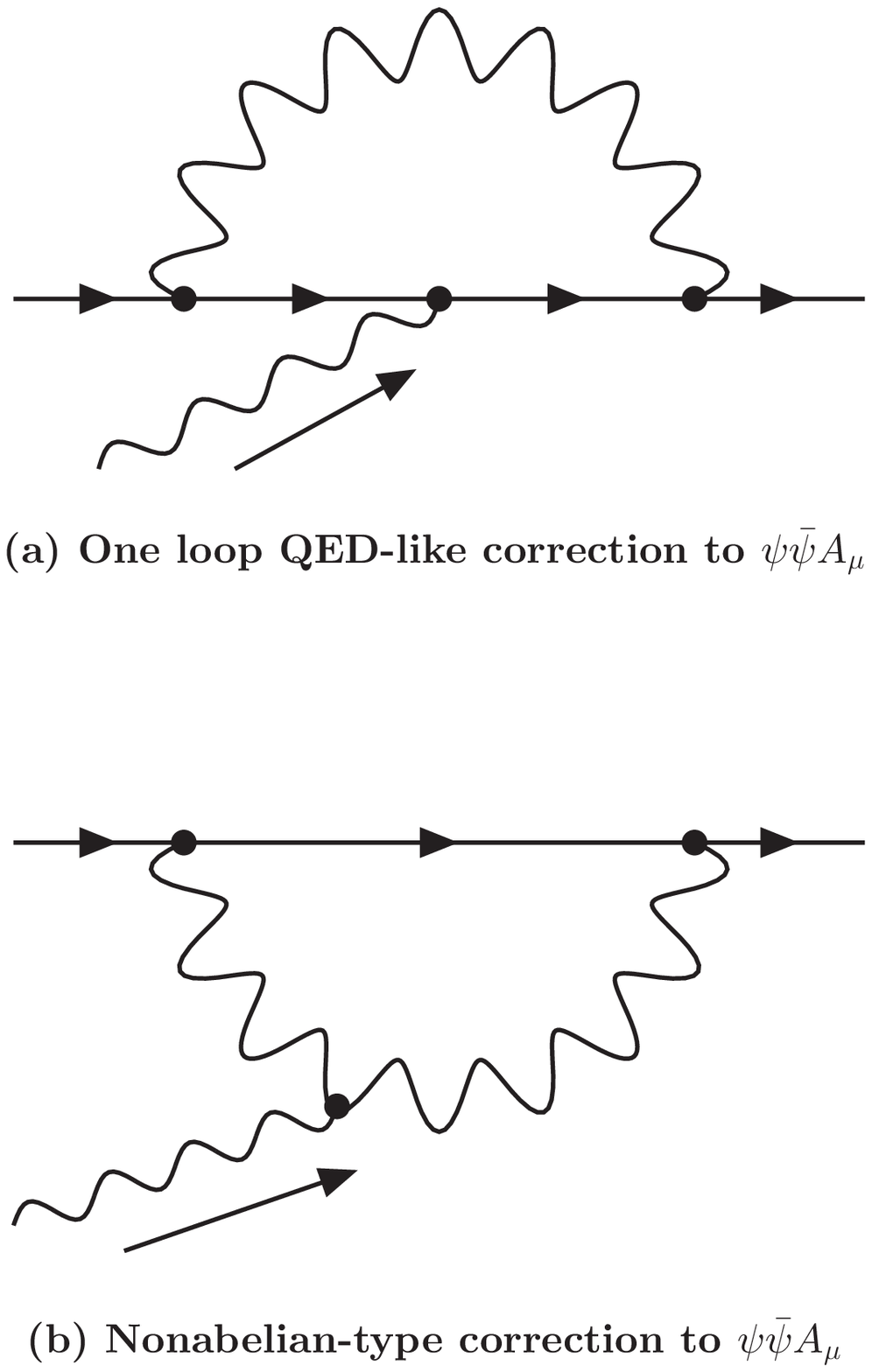}
\caption{One loop correction to $\psi\bar{\psi}A_\mu$ vertex}
\label{fig:g}
\end{center}
\end{figure}

%%%%%%%%%%%%%%%%%%%%%%%%%%%%%%%%%%%%%%%%%%%%%%%%%%%%%%%%%%%%%%%%%%%%%%%%%

The first diagram is similar to, what can also be found in the usual QED however, the second
diagram is completely new.
Now we proceed to write the analaytic expression of the first diagram,

\begin{eqnarray}
\Gamma^{(a)}_\mu & = &i(-ie)^2 \expp \int \frac{d^4k}{(2\pi)^4} 
e^{-ik\times q}
\frac{\gamma_\sigma}{k^2- \mg^2+i\epsilon}
\frac{\dir{p'}-\dir{k}+m}{(p'-k)^2-m^2+i\epsilon} \gamma_{\mu} 
\nonumber \\ & & 
\frac{\dir{p}-\dir{k}+m}{(p-k)^2-m^2+i\epsilon}\gamma^\sigma\ ,
\end{eqnarray}
where $m_{\gamma}$ is a photon mass accounting for the IR divergences.

Using the on mass shell conditions the numerator in the above expression may be written as 

\begin{equation}
4\Big\{\gamma_\mu \big[(p'-k)(p-k)-\frac{k^2}{2}\big]+\dir{k}(p'+p-k)_\mu - mk_\mu\Big\}.
\end{equation}

To compute the above integral, we replace the propagators by the Schwinger parameters:
\begin{equation}
\frac{i}{p^2-m^2+i\epsilon} = \int_0^\infty d\alpha e^{i\al(p^2-m^2+i\epsilon)}
\end{equation}
and use the auxiliary integral,
$$
\int
\frac{d^4k}{(2\pi)^4}e^{ik.(z-\tilde{q})}\frac{1}{(k^2-m_{\gamma}^2)(k^2-2p'.k)(k^2-2p.k)}= 
$$
%\nonumber \\
\begin{equation}
= \frac{1}{(4\pi)^2}\intalp \frac{exp\Big[\frac{-i}{a}\left(
(\aaalp)\al{1}m_{\gamma}^2+\big(\frac{(z-\tilde{q})}{2}-\al{2}p'-\al{3}p\big)^2\right)\Big]}{(\aaalp)^2}\
,
\end{equation}
where 
\begin{equation}
\tilde{q}_\nu = \theta_{\mu \nu}q^\mu\ , 
\end{equation}
and $\ \ \ a=\aaalp$.
The introduction of the factor $e^{ik.z}$ in the integrand allows us to obtain the required 
expression in the numerator of $\Gamma_\mu$ by differentiation \cite{itzykson}. 
After symmetrization in $\al{2}$ and $\al{3}$ we obtain,
\begin{eqnarray}
\Gamma_\mu^{(a)} & = & \frac{\alpha}{i\pi}\expp \intalp 
\frac{e^{\frac{i}{a\Lambda^2_{eff}}}}{a^3} 
exp{\Big[\frac{-i}{a}(\alpha_1 a
m_{\gamma}^2+(\alpha_2+\alpha_3)^2m^2-\alpha_2\alpha_3 q^2)\Big]}
\nonumber \\
& & e^{\frac{-i(\al{2}+\al{3})p.\tilde{q}}{a}}\ \Bigg\{\gamma_\mu \Big[ap'.p-
\frac{(\al{2}+\al{3})(p+p')^2}{2}+ \frac{i}{2} 
+\frac{(m^2(\al{2}+\al{3})^2-\al{2}\al{3}q^2)}{2a}\Big] 
\nonumber \\
&+& 
\frac{m}{2a}(p'+p)_\mu \alpha_1(\al{2}+\al{3})-
%\nonumber \\ 
\frac{\gamma_\mu}{4a}((p'+p)(\al{1}+\aaalp)-\frac{1}{2}\tilde{q}).\tilde{q} + \nonumber \\ 
& & +\frac{2m(\al{1}+2\al{2}+2\al{3})\tilde{q}_{\mu}+((p'+p)(2\al{1}+\al{2}+\al{3})-\tilde{q})_\mu 
\gamma.\tilde{q} }{4a}\Bigg\}\ ,
\end{eqnarray}
where we have inserted a UV regulator, $exp{\big(\frac{i}{a\Lambda^2}\big)}$, with the
following notation
\footnote{We should note that due to anti-symmetry of $\theta_{\mu\nu}$ ,  $\tilde{q}.\tilde{q}$ is negative
valued.}, 
\begin{eqnarray}
\Lambda^2_{eff} & = &\frac{1}{\Lambda^{-2}-\frac{\tilde{q}.\tilde{q}}{4}}\ .
\end{eqnarray}

After the Wick rotation, i.e $\al{i}\to\frac{\al{i}}{i}$, with the help of the following identity,

\begin{equation}
1=\int_0^\infty d\rho \delta(\rho-\sum \al{i})\ ,
\end{equation}
and rescaling $\al{i}\to\rho\al{i}$,
%$1=\int_0^\infty \frac{d\rho}{\rho} \delta(1-\sum \al{i})$,
(3.6) is obtained to be
%\begin{eqnarray}
%\Gamma_\mu^{(a)} & = & \frac{-\alpha}{\pi} \expp \inta e^{\frac{-i}{2}(\al{2}+\al{3})p.\tilde{q}} 
%\times \nonumber \\
%& & \times \Bigg\{\int^\infty_0 d\rho \bigg[\gamma_\mu\bigg(p'.p-
%\frac{(\al{2}+\al{3})(p'+p)^2}{2} 
%+ \frac{m^2}{2}(\al{2}+\al{3})^2 \nonumber \\
%& & -\frac{1}{2}\al{2}\al{3}q^2\bigg) + \frac{m}{2}\al{1}(\al{2}+\al{3})(p'+p)_\mu\bigg] -
%\nonumber \\ 
%& & -i \int_0^\infty \frac{d\rho}{\rho} 
%\bigg[\frac{\gamma_\mu}{2i}+\frac{\gamma_\mu}{4}((2-\al{2}-\al{3}))(p'+p).\tilde{q}+ \nonumber \\
%& & +\frac{m}{2}(1+\al{2}+\al{3})\tilde{q}_\mu +\frac{(\al{1}+1)}{4}(p'+p)_\mu \gamma.\tilde{q}
%\bigg]+ \nonumber \\
%& & +\int^\infty_0 \frac{d\rho}{4\rho^2} \bigg[\frac{-\gamma_\mu\tilde{q}.\tilde{q}}{2}+
%\tilde{q}_\mu \gamma.\tilde{q}\bigg]\Bigg\} \times \nonumber \\
%& & \times exp\left(-\rho (\T)-\frac{1}{\rho\Lambda^2_{eff}}\right),
%\end{eqnarray}
%or,
\begin{eqnarray}
\Gamma_\mu^{(a)}& = & \frac{-\alpha}{\pi} \expp \inta e^{-i(\al{2}+\al{3})p.\tilde{q}} \times 
\big( A_\mu+\frac{B_\mu}{\rho}+\frac{C_\mu}{\rho^2}\big)
\nonumber \\
& & \times \int_0^\infty d\rho\;\;  
exp\left(-\rho (\T) - \frac{1}{\rho \Lambda^2_{eff}}\right)\ , 
%\nonumber \\ & & 
\end{eqnarray}
where,
\begin{eqnarray}
A_\mu &=&  \gamma_\mu\Big[p'.p- \frac{(\al{2}+\al{3})(p'+p)^2)}{2} 
+ \frac{1}{2}m^2(\al{2}+\al{3})^2-\frac{1}{2}\al{2}\al{3}q^2\Big]+ \nonumber \\
& & \frac{m}{2}\al{1}(\al{2}+\al{3})(p'+p)_\mu\ , 
\end{eqnarray}
\begin{equation}
i B_\mu  = \frac{\gamma_\mu}{2i}+\frac{\gamma_\mu p.\tilde{q}}{2}(2-\al{2}-\al{3})
+\frac{m}{2}(1+\al{2}+\al{3})\tilde{q}_\mu + 
%\nonumber \\ & &+ 
\frac{1}{4}(\al{1}+1)(p'+p)_\mu \gamma.\tilde{q}\ , 
\end{equation}
\begin{equation}
C_\mu  =  \frac{-\gamma_\mu\tilde{q}.\tilde{q}}{8}+ \frac{\tilde{q}_\mu\gamma.\tilde{q}}{4}\ .
\end{equation}
Performing the integral over $\rho$, we get 
\begin{eqnarray}
\Gamma_\mu^{(a)}& = & \frac{-\alpha}{\pi} \expp \inta e^{-i(\al{2}+\al{3})p.\tilde{q}} \nonumber \\
& & \left(\frac{2A_\mu K_1(2\sqrt{X})}{\sqrt{X}\Lambda_{eff}^2}+2B_\mu K_0(2\sqrt{X})
+2\sqrt{X}\Lambda_{eff}^2 C_\mu K_1(2\sqrt{X})\right)\ ,
\end{eqnarray}
where 
$$
X\equiv\frac{\T}{\Lambda^2_{eff}}\ ,
$$ 
and $K_0 , K_1$ are the modified Bessel functions of the
first and second type respectively. 

%%%%%%%%%%%%%%%%%%%%%%%%%%%%%%%%%%%%%%%%%%%%%%%%%%%%%%%%%%%%%%%%%%%%%%%%%%
%%%%%%%%%%%%%%%%%%%%%%%%%%%%%%%%%%%%%%%%%%%%%%%%%%%%%%%%%%%%%%%%%%%%%%%%%%
\vskip .5cm

Now we consider the contribution from the second diagram, the analytic expression of which reads as
$$
\Gamma_{b}^\mu  =  -ie^2\expp \int\frac{d^4k}{(2\pi)^4} 
\frac{\left(1-e^{ik.\tilde{q}}e^{ip'\times p}\right)}{(k^2-m^2)((p'-k)^2-\mg^2)((p-k)^2-\mg^2)}\times 
$$
%\nonumber \\ & & 
\begin{equation}
\times\Bigg\{\gamma_\nu(\dir{k}+m)\gamma_\rho
\left[g^{\mu \nu}(2p'-p-k)^\rho+g^{\nu \rho}(2k-p'-p)^\mu+
g^{\rho \mu}(2p-p'-k)^\nu\right]\Bigg\}\  .
\end{equation}
Using the gamma matrices algebra  and the on shell condition the numerator can be written as,
\begin{equation}
8mk_\mu - 2\dir{k}(p'+p+2k)_\mu+2\gamma_\mu(2p'.k+2p.k-k^2-3m^2)\ .
\end{equation}
At this point we notice that (3.15) can be separated into two parts, 
one containing the phase $e^{-ik.\tilde{q}}$, and the other not, i.e.
$\Gamma^{b}_\mu=\Gamma^{b1}_\mu+\Gamma^{b2}_\mu$. 
First we perform the part containing the phase, $\Gamma^{(b2)}_\mu$, then the other term,
$\Gamma^{(b1)}_\mu$, can be recovered easily. 

Following the same spirit of the previous calculation, the auxiliary integral
is introduced and then doing the Gaussian integral,

\begin{eqnarray}
& & \hspace{-1cm} 
%-ie^2\expp 
\int\frac{d^4k}{(2\pi)^4} \frac{e^{ik.(z+\tilde{q})}}{(k^2-m^2)((p'-k)^2-\mg^2)((p-k)^2-\mg^2)}  =  
\nonumber \\
& & \hspace{-1cm} 
\frac{1}{4\pi^2}
\intalp \frac{1}{a^2} \;
e^{\frac{-i}{a}\left(\frac{z^2}{4}-z.(\al{2}p'+\al{3}p-\frac{\tilde{q}}{2})\right)}
e^{\frac{i}{a}(\al{2}+\al{3})p.\tilde{q}}
\nonumber \\ & & \hspace{-1cm} 
\times 
exp\left[-i\left(m^2(\al{1}-\al{2}-\al{3})+\mg^2(\al{2}+\al{3})+\frac{1}{a}(m^2(\al{2}+\al{3})^2-\al{2}\al{3}q^2)
-\frac{1}{a\Lambda_{eff}^2}\right)\right]\ , \nonumber \\
\end{eqnarray}

Producing the numerator through  the derivatives over the auxiliary field, a Wick
rotation, rescaling the $\alpha_i$'s and finally inserting the identity $1=\int_0^\infty
\frac{d\rho}{\rho} \delta(1-\sum \al{i})$,  
the expression $\Gamma^{(b2)}_\mu$ is found to be

\begin{eqnarray}
\Gamma^{(b2)}_\mu & &  =   \frac{-\alpha\expp}{\pi}\inta
e^{i(\al{2}+\al{3})p.\tilde{q}} e^{-ip\times p'} \times
\int  d\rho
\nonumber \\  & & \hspace{-0.2cm} 
exp~\left[-\rho\left(\N\right) - \frac{1}{\rho\Lambda_{eff}^2}\right]
\nonumber \\ & &
\times\left(\tilde{A}_\mu + \frac{\tilde{B}_\mu}{\rho} + \frac{\tilde{C}_\mu}{\rho^2}\right)\ ,
\end{eqnarray}

where now, $\tilde{A}_\mu, \tilde{B}_\mu ~\textrm{and}~ \tilde{C}_\mu$ are,
\begin{eqnarray}
\tilde{A}^\mu & = & \frac{1}{2}
\gamma_\mu\left[(\aalp)(p'+p)^2-3m^2-m^2(\aalp)^2+\al{2}\al{3}q^2\right]+ \nonumber \\
& & +\frac{m\al{1}}{2}(\aalp)(p'+p)_\mu \ ,
\end{eqnarray}
\begin{equation}
i \tilde{B}_\mu  = \frac{3i}{2}\gamma_\mu+\frac{1}{2}(m\tilde{q}_\mu\ +
\gamma_\mu p.\tilde{q})(2-\al{2}-\al{3})- 
\frac{\gamma.\tilde{q}}{4}(p'+p)_\mu(1+\al{2}+\al{3})\ ,
\end{equation}
\begin{equation}
\tilde{C}_\mu  = 
\frac{\gamma.\tilde{q} \tilde{q_\mu}}{4} + \frac{\gamma_\mu\tilde{q}.\tilde{q}}{8}\ .
\end{equation}
Integration over $\rho$ leads to
\begin{eqnarray}
\Gamma_\mu^{(b2)} & = & \frac{-\alpha \expp}{\pi}\inta e^{i(\aalp)p.\tilde{q}} e^{ip'.p}\times \nonumber \\
& &\times \left(\frac{2\tilde{A}_\mu K_1(2\sqrt{Y})}{\sqrt{Y}\Lambda_{eff}^2} +
 2\tilde{B}_\mu K_0(\sqrt{Y})+ \tilde{C}_\mu \sqrt{Y}\Lambda_{eff}^2 K_1(2\sqrt{Y})\right)\ ,
\end{eqnarray}
with
\begin{equation}
Y  \equiv  \frac{\N}{\Lambda^2_{eff}}\ .
\end{equation}

The $\Gamma^{(b1)}_\mu$ term, can be easily recovered from the above expression by setting any
term proportional to $\theta^{\mu \nu}$ (i.e. $\tilde{q}^\mu$) to zero. Then we have,
\begin{equation}
\Gamma_\mu^{(b1)}  =  \frac{-\alpha \expp}{\pi}\inta 
%\times \nonumber \\ & &
\left(\frac{2\bar{A}_\mu K_1(2\sqrt{Z})}{\sqrt{Z}\Lambda^2} + 2\bar{B}_\mu K_0(2\sqrt{Z})\right), 
\end{equation}
while $Z$ is now,
\begin{eqnarray}
Z &\equiv & \frac{\N}{\Lambda^2}.
\end{eqnarray}
and,
\begin{eqnarray}
\bar{A}^\mu & = & \frac{1}{2}
\gamma_\mu\left((\aalp)(p'+p)^2-3m^2-m^2(\aalp)^2+\al{2}\al{3}q^2\right)+ \nonumber \\
& &\frac{m\al{1}}{2}(\aalp)(p'+p)_\mu \\
\bar{B}_\mu & = & \frac{3}{2} \gamma_\mu\ .
\end{eqnarray}

\subsection{Renormalization}

Now we try to look at the divergences appearing in our diagrams. In the case of usual QED we have a logarithmic
UV divergence, and the problem of IR divergence was fixed by introducing the photon a finite mass, 
$\mg$. Looking at our expressions of the form factors , we see that they contain $K_0$, and $K_1$, and both of the 
functions contain either $\frac{1}{\Lambda_{eff}^2}$ or $\frac{1}{\Lambda^2}$ in their arguments. Taking the 
high energy limit $\Lambda^2 \to \infty$, or the low energy limit $q \to 0$ simultaneously we see that all terms 
containing $K_1$ are finite, but there appear to be a logarithmic divergence due to $K_0$. 
%We conclude that our diagrams up to the one loop are UV free, but they 
Therefore, we recover the same logarithmic divergence of
usual QED when taking the IR limit. 
The noncommutative QED was shown to be renormalizable up to the one loop level by adding the 
relevant counter terms \cite{Haya}, so we can safely drop the singular parts in $K_0$, and keeping only the
finite parts.
Now taking the $\Lambda\to \infty$ limit, and dropping the divergent parts
the renormalized Gamma functions
can be written as,
$$
\Gamma_{\mu (UV-ren)}^{(a)}  = \frac{-\alpha \expp}{\pi} \inta e^{-i(\al{2}+\al{3})p.\tilde{q}} \times 
$$
\begin{equation}
\times \left(\frac{A_\mu}{\T} - 2B_\mu \gamma_{Euler} + \Lambda_{eff}^2C_\mu\right), 
\end{equation}
$$
\Gamma_{\mu~(UV-ren)}^{(b1)}  = \frac{-\alpha \expp}{\pi}\inta \times 
$$
\begin{equation}
\times \Bigg(\frac{\bar{A}_\mu}{\N} 
-2\bar{B}_\mu \gamma_{Euller}\Bigg),
\end{equation}
and
$$
\Gamma_{\mu~(UV-ren)}^{(b2)} =  \frac{-\alpha \expp}{\pi}\inta e^{i(\aalp)p.\tilde{q}} 
e^{-ip\times p'}\times 
$$
\begin{equation}
\Bigg(\frac{\tilde{A}_\mu}{\N} -2 \tilde{B}_\mu \gamma_{Euller}+ \tilde{C}_\mu \Lambda_{eff}^2\Bigg)\ .
\end{equation}
Finally  the full renormalized Gamma function for the electron-photon vertex
at the one loop level is
\begin{eqnarray}\label{Gammaren}
\Gamma_{\mu (UV-ren)} & = & \frac{-\alpha \expp}{\pi} \inta  \times \nonumber \\
& & \times \Bigg(\frac{A_\mu e^{-i(\aalp)p.\tilde{q}}}{\T}+ \nonumber \\
& & + \frac{\bar{A}_{\mu}(1 - e^{i(\aalp)p.\tilde{q}} e^{-ip\times p'})}{\N} - \nonumber \\
& & -2\gamma_{Euler}(B_\mu e^{-i(\aalp)p.\tilde{q}}+\bar{B}_\mu -\tilde{B}_\mu e^{i(\aalp)p.\tilde{q}}
e^{-ip\times p'}) 
\nonumber +\\
& & + \Lambda_{eff}^2(C_\mu e^{-i(\aalp)p.\tilde{q}} - \tilde{C}_\mu e^{i(\aalp)p.\tilde{q}} e^{-ip \times
p'})\Bigg)\ .
\end{eqnarray}
In equation (\ref{Gammaren}), we have given the UV - renormalized gamma function, but we can see that in the last
term we still
keep the cut-off in the expression, and this can be understood by the UV/IR mixing. In order this we will
discuss both limits applied to this term, first considering the UV limit, i.e.  
$\frac{1}{\Lambda^2} \ll \tilde{q}.\tilde{q}\ \ ( \Lambda^2_{eff} \sim
\frac{1}{\tilde{q}.\tilde{q}})$, 
we see that the term is finite, but when taking the IR limit first, i.e.
$\frac{1}{\Lambda^2} \gg \tilde{q}.\tilde{q}$ then
$\Lambda^2_{eff} \sim \frac{1}{\Lambda^2}\ $, so it may seem that this term will lead to an IR divergence.
However, $C_{\mu}$ terms contain two type of terms, both proportional to $\tilde{q}^2$, and since
${\Lambda^2}{\tilde{q}^2} \ll 1$, in the IR limit, this term will be totally irrelevant. 
One should note that order of taking the 
${\Lambda^2}\to\infty$ and $q\to 0$ is very important in our arguments; this is a generic feature of NCFT's
and  is called IR/UV mixing \cite{{seib},{Suss}}.
Hence the fully renormalized vertex function is,
\begin{eqnarray}
\Gamma_{\mu (ren)} & = & \frac{-\alpha \expp}{\pi} \inta  \times \nonumber \\
& & \times \Bigg(\frac{A_\mu e^{-i(\aalp)p.\tilde{q}}}{\T}+ \nonumber \\
& & + \frac{\bar{A}_{\mu}(1 - e^{i(\aalp)p.\tilde{q}} e^{-ip\times p'})}{\N} - \nonumber \\
& & -2\gamma_{Euler}(B_\mu e^{-i(\aalp)p.\tilde{q}}+\bar{B}_\mu -\tilde{B}_\mu e^{i(\aalp)p.\tilde{q}}
e^{-ip\times p'})\Bigg)\ .
\end{eqnarray}
For the moment we can see that $\Gamma^\mu$ can be written in the form,
\begin{eqnarray}
\Gamma^\mu  & = &  E_1\gamma^\mu + H_1(p'+p)^\mu + G_1\tilde{q}^\mu + E_2\gamma^\mu p.\tilde{q} 
+H_3(p'+p)^\mu\gamma.\tilde{q}.
\end{eqnarray}
Collecting the different cofficients of
$\gamma^\mu, (p'+p)^\mu, \tilde{q}^\mu,\gamma^\mu p.\tilde{q}^\mu,(p'+p)^\mu\gamma.
\tilde{q}$, we have
\begin{eqnarray}
E_1 & = & \frac{-\alpha \expp}{\pi} \inta \times 
\left(1 - e^{i(\aalp)p.\tilde{q}} e^{-ip\times p'}\right)
\nonumber \\
&\times & 
\Bigg\{\left[\frac{\left(2p'.p-(\al{2}+\al{3})(p'+p)^2)+m^2(\al{2}+\al{3})^2-\al{2}\al{3}q^2\right)}{2(\T)} 
+ \gamma_{Euller} \right]e^{-i(\aalp)p.\tilde{q}} 
\nonumber \\
&+& \left[\frac{\left((\aalp)(p'+p)^2-3m^2-m^2(\aalp)^2+\al{2}\al{3}q^2\right)}{\N} 
- \frac{3\gamma_{Euller}}{2}\right]\Bigg\}\ , 
\nonumber \\
H_1 & = & \frac{-\alpha \expp}{\pi} \inta  \times 
\nonumber \\
& & \times \Bigg\{\frac{m\al{1}(\aalp)e^{i(\al{2}+\al{3})p.\tilde{q}}}{\T} + 
\nonumber \\
& & + \frac{m\al{1}(\aalp)\left(1 - e^{i(\aalp)p.\tilde{q}}e^{-ip \times p'}\right)}{\N} \Bigg\}\ , 
\nonumber \\
G_1 & = & \frac{-\alpha \expp}{\pi} im\gamma_{Euller}\inta \times 
\nonumber \\ 
& & \times \left((1+\al{2}+\al{3})e^{-i(\al{2}+\al{3})p.\tilde{q}} - (2-\al{2}-\al{3})
e^{i(\aalp)p.\tilde{q}}e^{-ip\times p'}\right)\ , \nonumber 
\end{eqnarray}
$$
E_2  =  \frac{-\alpha \expp}{\pi} (i\gamma_{Euller})\inta 
\left(1 -e^{-ip \times p'}\right)(2-\al{2}-\al{3}) e^{-i(\al{2}+\al{3})p.\tilde{q}}
%e^{-i(\al{2}+\al{3})p.\tilde{q}}
$$
\begin{eqnarray}
H_3 & = & \frac{-\alpha \expp}{\pi} \frac{i\gamma_{Euller}}{2} \inta \times 
\nonumber \\
& & \left((2-\al{2}-\al{3})e^{-i(\al{2}+\al{3})p.\tilde{q}} + (1+\al{2}+\al{3})
e^{i(\aalp)p.\tilde{q}}e^{-ip\times p'}\right)\ .
\end{eqnarray}

In the next section we will study the physical significance of these factors.

The only comment we would like to make in this part is the appearance of extra factors of
$e^{-ip\times p'}$ in the above expressions. More precisely, our vertex function in the one loop
level contains two different terms, one proportional to $e^{\frac{i}{2}p\times p'}$ and the other
proportional to $e^{-\frac{i}{2}p\times p'}$. This can be understood if we note that our thoery
is not CP invariant and if we change the arrow of time in our diagrams we will find the same
expression, but we also should change $\theta$ to $-\theta\ $ \cite{CPT} .

\newsection{The electron-photon vertex function}

In this section we give a formal discussion of the electron-photon 
vertex structure found in NCQED.  So, first we  review the similar argument in the usual QED. The
physical meaning of the different terms appearing in our vertex from the calculation in 
the previous section will be discussed later. 

%%%%%%%%%%%%%%%%%%%%%%%%%%%%%%%%%%%%%%%%%%%%%%%%%%%%%%%%%%%%%%%%

\subsection{Electron-photon vertex in usual QED}

In this section we briefly present the structure of the electron-photon vertex in the usual QED.
Considering the radiative corrections to the vertex, it can be written in the following a diagrammatic 
form

%%%%%%%%%%%%%%%%%%%%%%%%%%%%%%%%%%%%%%%%%%%%%
\begin{figure}[h]
\begin{center}
\includegraphics[scale=.8]{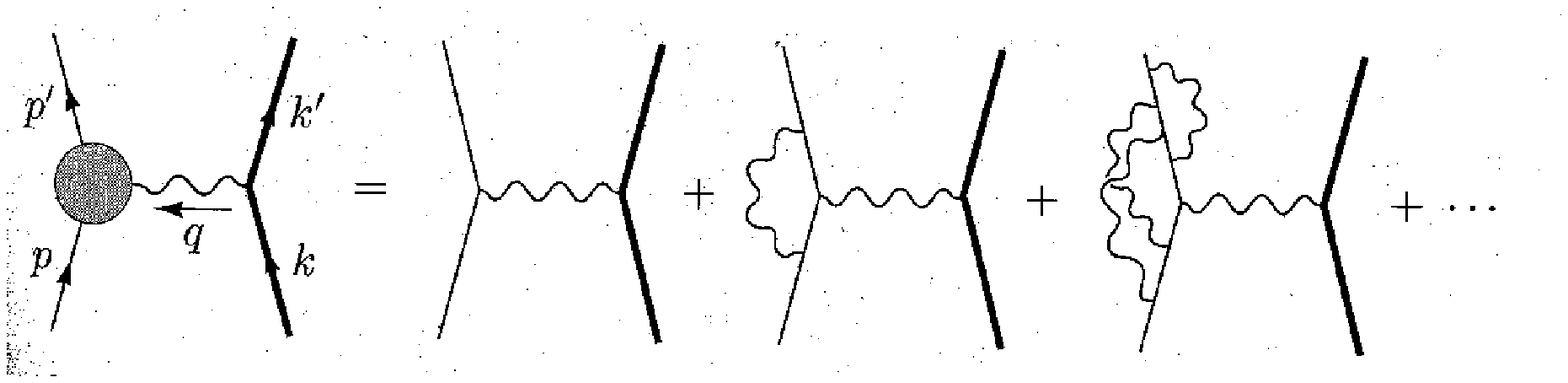}
\label{ihab2}
\end{center}
\end{figure}
%%%%%%%%%%%%%%%%%%%%%%%%%%%%%%%%%%%%%%%%%%%%% 

where the shaded circle can be expressed as

\vspace{1cm}

\begin{equation}
\includegraphics[bbllx=70,bblly=380,bburx=600,bbury=500,width=2cm]{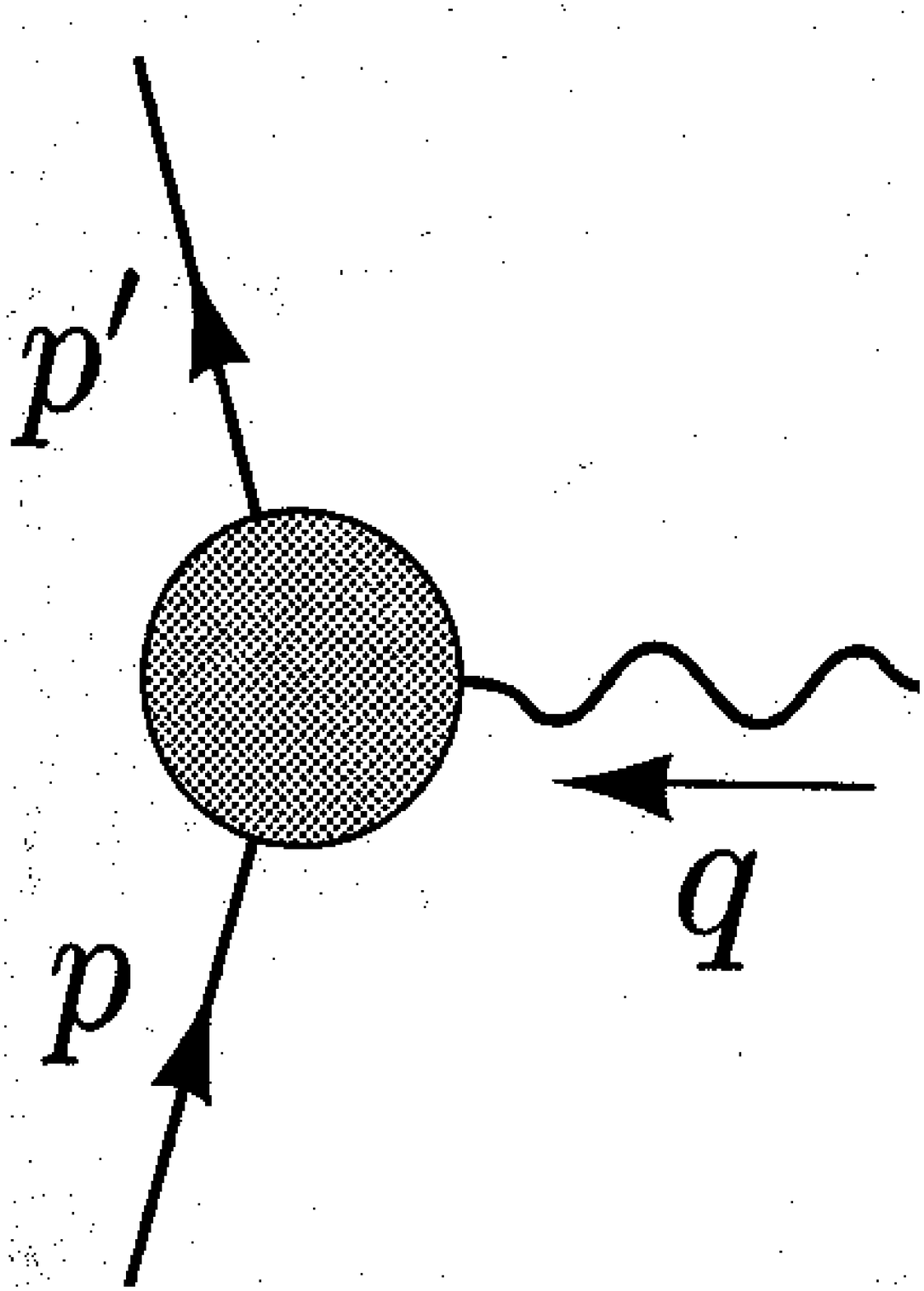} 
= -ie\bar{U}(p')\Gamma^\mu(p',p)U(p)\tilde{A}_\mu(q).
\end{equation}

\vspace{1cm}

We observe that in the leading order of perturbation $\Gamma^\mu \to \gamma^\mu$. 

The list of  available vectors and scalars is 
only consisting of $(p_\mu ~,p'_\mu ~,\gamma_\mu, q^2, ~m,~e)\ $.
So, the most general form of the vertex function can be written as
\footnote {We have not included $\gamma^5$ because of the parity conservation.}
\begin{equation}
\Gamma^\mu = A\gamma^\mu + B(p'+p)^\mu + C (p'-p)^\mu\ ,
\end{equation}  
where A, B, C can be any scalars formed out of  previous list. 
Appealing to Ward identity i.e. $q^\mu \Gamma_\mu = 0$\ ,
we get
\begin{equation}
0 = A\dir{q} + B(p'+p).q + C(p'-p).q\ .
\end{equation}
Since, $\bar{U}(p')\dir{q}U(p) = 0~~, (p'+p).q = 0~~\textrm{and}\;\;\; (p'-p).q \not\!\!{=} 0$, 
we conclude that to satisfy the Ward identity, $C = 0$.
Therefore, $\Gamma^\mu$ for QED takes the following form,
\begin{equation}
\Gamma^\mu = A\gamma^\mu + B(p'+p)^\mu\ ,
\end{equation}
using the Gordon identity we can write the above expression as,
\begin{equation}
\Gamma^\mu = F_1(q^2)\gamma^{\mu} + \frac{i\sigma^{\mu
\nu}q_\nu}{2m}F_2(q^2)\ .
\end{equation}
$F_1(q^2)~~\textrm{and}~F_2(q^2)$ which usually are known as form factors, 
are functions of m, e and $q^2$ depending on the diagram.
One can see that at the tree level we have $\Gamma^\mu \to \gamma^\mu$, hence 
$F_1(q^2) = 1~\textrm{and}\;\; F_2(q^2)=0$.

To give an interpretation to the above form factors we will couple this vertex to external slowly varying 
electric and magnetic fields at the tree level.

%%%%%%%%%%%%%%%%%%%%%%%%%%%%%%%%%%%%%%%%%%%%%%%%%%%%%%%%%%%%%%%%%%%%%%%%%%%
\vskip .5cm
{\bf The electron electric charge}
%\newline
\vskip .2cm
Let us  consider the Coulomb scattering cross section, from a slowly varying electric field, 
i.e. we can take the limit $q^2 = 0$. The corresponding  scattering amplitude can be written as
\begin{equation}
-iM = -ieF_1(0)\tilde{\Phi}(q)\xi^{\dagger}\xi\ ,
\end{equation}
where we have used the non-relativistic approximation for the spinor field,
\begin{equation}\label{nonrel}
U(p) = \frac{1}{\sqrt{2}}
\left( \begin{array}{c} \left(1-\frac{\vec{p}.\vec{\sigma}}{2m}\right)\xi \\
\left(1+\frac{\vec{p}.\vec{\sigma}}{2m}\right)\xi
\end{array}\right).
\end{equation}
In the Born approximation $V(x)=eF_1(0)\Phi(x)$, hence we can identify $F_1(0)$ as the electric charge in units of
e, and since $F_1(0) = 1$ at the tree level, radiative corrections to $F_1(q^2)$ 
should vanish at $q^2 = 0$.

%%%%%%%%%%%%%%%%%%%%%%%%%%%%%%%%%%%%%%%%%%%%%%%%%%%%%%%%%%%%%%%%%%%%%%%%%%%
\vskip .5cm
{\bf The electron magnetic moment}
%\newline
\vskip .2cm
We now repeat the analysis for an electron scattering from a static vector potential. 
The amplitude of scattering from this field is
\begin{equation}
iM = ie \left[\bar{U}(p')\left(\gamma^iF_1+\frac{i\sigma^{i\nu}}{2m}q_\nu F_2\right)U(p)\right]
\tilde{A}^i(q)\ .
\end{equation}
Inserting the non-relativistic expansion for the spinors (\ref{nonrel}), and keeping terms 
first order in momenta we obtain,
\begin{equation}
iM= -ie \xi^{\dagger}\left(\frac{-\sigma_k}{2m}[F_1(0)+F_2(0)]\right)\xi\tilde{B}^k(q)\ ,
\end{equation}
where
\begin{equation}
\tilde{B}^k(q)=-i\epsilon^{ijk}q_i\tilde{A}_j(q)\ ,
\end{equation}
is the Fourier transform of the magnetic field produced by $A(x)$.

Again we interpret ~ $M$ ~as the Born approximation to the scattering of the electron from a potential.
The potential is just that of a magnetic moment interaction,
$V(x)=-\langle\mu\rangle .B(x)$,
where,
\begin{equation}
\langle \vec{\mu} \rangle = \frac{e}{m}[F_1(0)+F_2(0)]{\xi}^{\dagger} \frac{\vec{\sigma}}{2}\xi\ .
\end{equation}
This expression for the magnetic moment of the electron can be rewritten in the standard form
\begin{equation}
\vec{\mu} = g \left(\frac{e}{2m}\right)\vec{S}\ ,
\end{equation}
where $\vec{S}$ is the electron spin. The coefficient $g$ is
\begin{equation}
g = 2[F_1(0)+F_2(0)] = 2 + 2F_2(0)\ .
\end{equation}
Along our previous argument for the leading order of $F_1$ and $F_2$, we see that to the leading order
(classical level) the magnetic moment of a Dirac particle is 2.
%%%%%%%%%%%%%%%%%%%%%%%%%%%%%%%%%%%%%%%%%%%%%%%%%%%

\subsection{Dipole moment at the tree level in NCQED}

The calculation of the vertex function at the tree level goes exactly as in the case of usual QED, 
except for the extra phase $\expp$ appearing due to the star product, i.e.
\begin{eqnarray}
\Gamma^\mu & = & \expp \gamma^\mu\ , \nonumber \\
& = & e^{\frac{i}{2}p.\tilde{q}} \gamma^\mu\ .
\end{eqnarray}
Using a power expansion of the exponential and keeping only the first two terms, we see that the 
first term in the expansion which is $\theta$ independent gives rise to the usual result in QED, 
i.e. we get the same electric charge and magnetic dipole moment. 
The second term which is 
proportional to $\theta$ will give rise to an {\it electric dipole moment} and couples to the 
external electric field, $E$,  as $\langle P \rangle.E$, where
\begin{equation}\label{elecdip}
\langle P_i \rangle = \frac{1}{2} e \tilde{p}_i\ = \frac{1}{2} e\theta_{ji}p_j \ .
\end{equation}
This term will also contribute to higher-pole moments when
coupled to an external electro-magnetic field.

%%%%%%%%%%%%%%%%%%%%%%%%%%%%%%%%%%%%%%%%%%%%%%%%%%%

\subsection{The electron-photon vertex structure in NCQED}

In the above we discussed the structure of the electron-photon vertex, in the case of usual QED. The list
of vectors and scaler appearing in the vertex function was restricted to 
$
\left(\gamma^\mu, (p'+p)^\mu, q^2, m, e \right)\ .
$
In the case of NCQED, and due to the presence of $\theta^{\mu \nu}$, our previous list of independent
vectors and scalars will be extended to include  three other scalars 
$$
\left(q^2, m, e, \tilde{q}.\tilde{q},\gamma.\tilde{q} , ~p.\tilde{q}\right)\ ,$$
and the list for the vectors will be
$$\left(\gamma^\mu, (p'+p)^\mu ,~q^{\mu},  ~\tilde{q}^\mu \right)\ .$$ 
The most general structure of the vertex function which is compatible with the Ward
identity, is
\begin{eqnarray}
\Gamma^\mu  & = &  E\gamma^\mu + H(p'+p)^\mu + G\tilde{q}^\mu\ ,
\end{eqnarray}
where E, F and G are scalars formed from the list of our previous scalars,  except for $\tilde{q}.\tilde{q}$,
which as discussed in the previous section will not appear in the first loop.
Up to the one loop approximation the coefficients E, F and G may be expanded 
to be written as functions of $q^2, p.\tilde{q}~~\textrm{and}~~ \gamma.\tilde{q}$. 
The form factors $F_1(q^2)~~\textrm{and}~~F_2(q^2)$ can be picked directly 
from $E_1~~\textrm{and}~~H_1$ using the Gordon identity. From the results of the previous calculations 
one can see that the only non-zero coefficients are $E_1,~E_2,~H_1,~H_3~\textrm{and}~G_1$, i.e. 
\begin{eqnarray}
\Gamma^\mu  & = &  E_1\gamma^\mu + H_1(p'+p)^\mu + G_1\tilde{q}^\mu + E_2\gamma^\mu p.\tilde{q} 
+H_3(p'+p)^\mu\gamma.\tilde{q}.
\end{eqnarray}
So what is left now is to give
a physical interpretation to those coefficients proportional to $\theta$. We will use the non-relativistic
limit, 
to compute $\bar{U}(p')\Gamma^\mu U(p)$.  Keeping terms up to second order of momentum we find,
\begin{itemize}
\item  $G_1$, coefficients of $\tilde{q}_\mu$:

This term will give a contribution to the magnetic moment. The corresponding effective interaction 
potential with the external magnetic filed is 
$V(x)=-\langle\mu\rangle .B(x)$, where 
\begin{equation}
\langle\vec{\mu} \rangle = \frac{G_1}{i}\vec{\theta}\ ,\;\;\;\;\; 
{\theta}_i\equiv\ \epsilon_{ijk}\theta_{jk}\ .  
\end{equation}

As we see this magnetic moment does not depend on spin.

\item $E_2$, coefficients of $\gamma^\mu p.\tilde{q}$:

This term will vanish in the non-relativistic limit, but 
it will give contribution to higher moments when higher 
orders of momentum are considered.

\item  $H_3$, coefficients of $(p'+p)^\mu \gamma.\tilde{q}$:

This term will give rise to an electric dipole moment of the form 
$$
\langle P \rangle_i = 2iH_3 \tilde{p}_i\ . 
$$ 
%and there will also be a term of the form
%$(-2i\epsilon^{ijk}H_3 \sigma_k \tilde{q}_i E_j)$.
\end{itemize}

Since we are going to work in the low momentum limit, we can use the series expansion of
$e^{-i(\al{2}+\al{3})p.\tilde{q}} , e^{i(\aalp)p.\tilde{q}}~~ \textrm{and} ~~ e^{ip'\times p}$
 and  keep only the first term of the expansion. In this limit  
\begin{eqnarray}
E_1 & = & \frac{-\alpha \expp}{\pi} \inta \times \nonumber \\
& & \times \left[\frac{\left(2p'.p-(\al{2}+\al{3})(p'+p)^2)+m^2(\al{2}+\al{3})^2-\al{2}\al{3}q^2\right)}{2(\T)} 
+ \gamma_{Euller} \right]\ , \nonumber \\
H_1 & = & \frac{-\alpha \expp}{\pi} \inta \frac{m\al{1}(\aalp)}{\T}\ , \nonumber \\
G_1 & = & \frac{-\alpha \expp}{\pi} im\gamma_{Euller}\inta (\aalp-1)\ , \nonumber \\
H_3 & = & \frac{-\alpha \expp}{\pi} \frac{3i\gamma_{Euller}}{2} \inta\ .
\end{eqnarray}

Now we can apply the Gordon identity to $E_1$ and $H_1$, and identify the form factors 
$F_1(q^2) ~\textrm{and} ~ F_2(q^2)$ directly

$$
F_1(q^2)  =  \frac{-\alpha \expp}{\pi} \inta \times 
$$ 
\begin{equation}\label{F1}
\times \left[ \frac{2m^2(1-\al{2}-\al{3})-q^2(1-\al{2}-\al{3})-m^2(\aalp)^2-\al{2}\al{3}q^2}{2(\T)} +
\gamma_{Euller}\right]\ , 
\end{equation}

and 
\begin{equation}\label{F2}
F_2(q^2)  =  \frac{\alpha \expp}{\pi} \inta 
\frac{m^2\al{1}(\aalp)}{\T}\ .
\end{equation}
As it is seen, apart from the pre-factor, $e^{\frac{i}{2} p\times p'}$, the (\ref{F1}), (\ref{F2}) expressions
are exactly the same as the usual QED.

The integration over the Schwinger parameters in $G_1, ~\textrm{and}~ H_3$ terms can be easily 
performed,
\begin{eqnarray}
G_1 & = & \frac{\alpha \expp}{\pi} \left(\frac{im\gamma_{Euller}}{6}\right), \nonumber \\
H_3 & = & \frac{-\alpha \expp}{\pi} \left(\frac{3i\gamma_{Euller}}{4}\right)\ .
\end{eqnarray}

%\begin{eqnarray}
%\langle P \rangle_i & = & 3m\alpha\gamma_{Euller}\tilde{p}_i e^{\frac{1}{2}p' \times p}, \nonumber \\
%\langle \mu \rangle_i & = & m^2\alpha\gamma_{Euller} \theta_k e^{\frac{1}{2}p' \times p}\ .
%\end{eqnarray}

Altogether,  we can write the interaction of a noncommutative  electron and an external electromagnetic 
field in the form,
\begin{equation}
V(x) = e\Phi + \langle \mu \rangle .B + \langle P \rangle .E\ ,
\end{equation}
where the first term is just a Coulomb potential, the second is the magnetic 
dipole moment and the last term 
is the electric dipole moment. 
%Now we consider writting this as a perturbation 
%expansion up to one loop,
%\begin{equation}
%V(x) = e\Phi + \langle \mu \rangle^{(0)} .B + \langle \mu \rangle^{(1)} .B 
%\langle P \rangle^{(0)} .E
%+ \langle P \rangle^{(1)} .E ,
%\end{equation}
The  coefficients $\langle \mu \rangle$ and $\langle P \rangle$, in the low momentum approximation, and
up to first loop are
\begin{eqnarray}
\langle \vec{\mu} \rangle & = & \frac{e}{m}\bigg[(F_1(0) + F_2(0))\vec{S}+ \frac{\alpha 
\gamma_{Euller}}{6\pi}\  m^2\vec{\theta}\bigg]\ ,\\
\langle \vec{P} \rangle & = & \frac{1}{4}e(\vec{\theta}\times\vec{p})\; (1+3\alpha
\gamma_{Euller})
\end{eqnarray}

\newsection{Concluding remarks}

In this work we have discussed some aspects of NCQED. First we introduced the theory by giving its action, and
the corresponding basic Feynman graphs. 
We argued that the photon itself, similar to the moving noncommutative electron, even at classical level, 
shows some electric dipole effect. However, the dipole moment of the fermions compared to that of photon, is
less by a factor of one half. This is due to the fact that in the NCYM action, the noncommutative effects
appear as the Moyal-bracket, while in the fermionic sector it is just a star product.

Calculating the one loop contributions to the electron-photon vertex, we studied the 
electro-magnetic dipole moments of the electron. We showed that magnetic dipole
moment of electron has now two parts, one spin dependent which will not receive
any further corrections due to the noncommutativity, and the other spin independent,
being proportional to $\theta$. In addition we also found  the one loop
contributions to the electric dipole moment.

Here we only studied the loop effects coming form the interaction vertex correction, however
there are many other interesting phenomena coming from the vacuum polarization and corrections to 
propagators such as, Lamb shift, or the pair production threshold, which we postpone them to the future
studies.

%\thispagestyle{empty}
%\null
\vskip 1cm
%\begin{center}
{\large\bf{Acknowledgements}}
%\end{center}

One of us, I.F. R. ,  would like to thank all the Professors of ICTP HEP Diploma Course,
Ms. Concetta Mosca for her endless help throughout the year, UNESCO, the IAEA and
Professor M.A. Virasoro, Director of ICTP, for their kind hospitality at the Centre
during the Diploma course programme.
M.M. Sh-J. would like to thank T. Krajewski for helpful comments. 

This research was partly supported by the EC contract no. ERBFMRX-CT 96-0090.

\appendix

\newsection{Some useful identities in *-product calculus}
\label{A}  
Let $f, g$ be two arbitrary functions on non-commutative $R^d$:
$$
f(x)=\int f(k) e^{ik.x}d^dk, \;\;\;\;
g(x)=\int g(k) e^{ik.x}d^dk\ .
$$
Then
$$
(f*g)(x)=\int f(k) g(l)e^{-ik\theta l/2} e^{i(k+l).x}d^dkd^dl\ ,
$$
where $k\theta l=k^{\mu}\theta_{\mu\nu}l^{\nu}$. From the above relation
it is straightforward to see:

1) $g*f=f*g|_{\theta\rightarrow -\theta}$, and hence
$\{f,g\}_{M.B.}=f*g|_{\theta} -f*g|_{-\theta}$ .

2) $\int (f*g) (x) d^dx=\int (g*f) (x) d^dx=\int fg (x) d^dx$ .

3) If we denote  complex conjugation by $c.c.$, then

$(f*g)^{c.c.}=g^{c.c.}*f^{c.c.}$ . 

If $h$ is another arbitrary function:

4) $(f*g)*h=f*(g*h)\equiv f*g*h$ .

5)$\int (f*g*h)(x) d^dx=\int (h*f*g)(x) d^dx=\int (g*h*f)(x) d^dx$ .

6) $(f*g*h)|_{\theta}= (h*g*f)|_{-\theta}$ .

In other words the integration on the space coordinates, $x$, has the
cyclic
property, and it has all the properties of the $Tr$ in the matrix calculus.

From 2) we learn that the kinetic part of the actions (which are quadratic
in fields) is the same as their commutative version. So the free field
propagators in commutative and nnoncommutative spaces are the same.

\end{document}